\documentclass[aps,prl,twocolumn,showpacs,superscriptaddress,groupedaddress]{revtex4}  
\usepackage{graphicx}  
\usepackage{dcolumn}   
\usepackage{bm}        
\usepackage{amssymb}   
\usepackage{amsmath}
\usepackage{float}
\usepackage{xcolor}

\begin{document}



\title{Unified Viscous-to-inertial Scaling in Liquid Droplet Coalescence}

\author{Xi Xia}
\affiliation{
Department of Mechanical Engineering, The Hong Kong Polytechnic University, Hung Hom, Kowloon, Hong Kong}
\affiliation{
School of Mechanical Engineering, Shanghai Jiao Tong University, Minhang, Shanghai, P. R. China}
\author{Chengming He}
\author{Peng Zhang}%
\email{pengzhang.zhang@polyu.edu.hk}
\affiliation{
Department of Mechanical Engineering, The Hong Kong Polytechnic University, Hung Hom, Kowloon, Hong Kong}
\date{\today}

\begin{abstract}
This letter presents a theory on the coalescence of two spherical liquid droplets that are initially stationary. The evolution of the radius of a liquid neck formed upon coalescence was formulated as an initial value problem and then solved to yield an exact solution without free parameters, with its two asymptotic approximations reproducing the well-known scaling relations in the viscous and inertial regimes. The viscous-to-inertial crossover observed by Paulsen \textit{et al.} [Phys. Rev. Lett. {\bf 106}, 114501 (2011)] is also recovered by the theory, rendering the collapse of data of different viscosities onto a single curve.
\end{abstract}

\maketitle


Droplet Coalescence \cite{EggersJ:99a,AartsDGAL:05a,Thoroddsen:05a,PaulsenJD:11a,Xia:17a} is a ubiquitous phenomenon involving impact or contact of dispersed two-phase flows \cite{YarinAL:06a,ZhangP:11a,ThoravalMJ:12a,TranT:13a,KavehpourHP:15a,ZhangP:16a,Xia:19a}. Among the various relevant problems, the initial coalescence of two liquid droplets has been of core interest. The first quantitative analysis of sphere coalescence was provided by Frenkel \cite{FrenkelJJ:45a} based on the assumption of internal Stokes flow; however, the result was commented as ``misleading" by Hopper \cite{HopperRW:93a}, who gave an analytical solution for the coalescence of two cylindrical droplets of radius $R_0$ for viscous sintering. His studies \cite{HopperRW:90a,HopperRW:92a} show that the time evolution of the radius $R$ of the neck (or bridge) between the droplets approximately satisfies $t \sim -R/\ln{R^{\ast}}$ where $R^{\ast}=R/R_0$. Later, Eggers \textit{et al.} \cite{EggersJ:99a} considered the three-dimensional coalescence and attained $R^{\ast} \sim -t^{\ast} \ln{t^{\ast}}$ for $R^{\ast} < 0.03$, where $t^{\ast} = t/\tau_v$ ($\tau_v=\mu R_0/\sigma$ with $\mu$ and $\sigma$ being the dynamic viscosity of the liquid and the surface tension coefficient, respectively). For larger $R^{\ast}$, they \cite{EggersJ:99a,DucheminL:03a} argued that the neck flow goes beyond the Stokes regime to the inertial (or inviscid) regime, and further arrived at the $1/2$ power-law scaling, $R^{\ast} \sim (t/\tau_i)^{1/2}$ with the time scale being $\tau_i=(\rho R_0^3/\sigma)^{1/2}$, where $\rho$ is the liquid density.

Recent advances in the high-speed digital imaging \cite{AartsDGAL:05a,Thoroddsen:05a,YaoW:05a}, state-of-art probing techniques \cite{CaseSC:08a,FezzaaK:08a,PaulsenJD:11a}, and numerical simulation \cite{SprittlesJE:12a,SprittlesJE:14a} enabled researchers to scrutinize the early stages of drop coalescence when $R^{\ast} \ll 1$. As a result, the $1/2$ power-law scaling was confirmed by many experimental \cite{WuM:04a,AartsDGAL:05a,Thoroddsen:05a,BurtonJC:07a,FezzaaK:08a,CaseSC:09a} and numerical \cite{DucheminL:03a,EiswirthRT:12a,PothierJC:12a,GrossM:13a,SprittlesJE:12a} studies. The same scaling was also observed for droplet coalescence on substrate \cite{StoneHA:06a,EddiA:12a,EddiA:13a}. However, the experiments of Aarts \textit{et al.} \cite{AartsDGAL:05a} and Thoroddsen \textit{et al.} \cite{Thoroddsen:05a} indicate that the viscous regime is well predicted by the linear scaling of $R^{\ast} \sim t^{\ast}$, noting that most of their data were in the $R^{\ast} > 0.03$ range. This linear correlation was also corroborated by other studies \cite{YaoW:05a,BurtonJC:07a,PaulsenJD:11a}.

More recently, research interests have been directed towards the crossover (or transition) between the viscous and inertial regimes. The first direct evidence of the crossover from $R^{\ast} \sim t^{\ast}$ to $R^{\ast} \sim (t^{\ast})^{1/2}$ was reported by Burton and Taborek \cite{BurtonJC:07a}. By equating the characteristic velocities from the two scaling laws, they derived the crossover length, $l_c \sim \mu(R_0/\rho \sigma)^{1/2}$, which was later confirmed by Paulsen \textit{et al.} \cite{PaulsenJD:11a,PaulsenJD:13a}, who further obtained the crossover time, $\tau_c \sim \mu^2(R_0/\rho \sigma^3)^{1/2}$. With these time and length scales, Paulsen \textit{et al.} applied a fitting curve, $(R/l_c)^{-1} \sim (t/\tau_c)^{-1}+(t/\tau_c)^{-1/2}$, to collapse the neck evolutions of distinct viscosities, which points to a universality in droplet coalescence. To theoretically explore this universality, we derived a scaling model for the viscous-to-inertial combined coalescence regime \cite{Xia:18a}, with two scaling constants determined by fitting experimental data. 

\begin{figure}
\begin{center}
\includegraphics[scale=0.3]{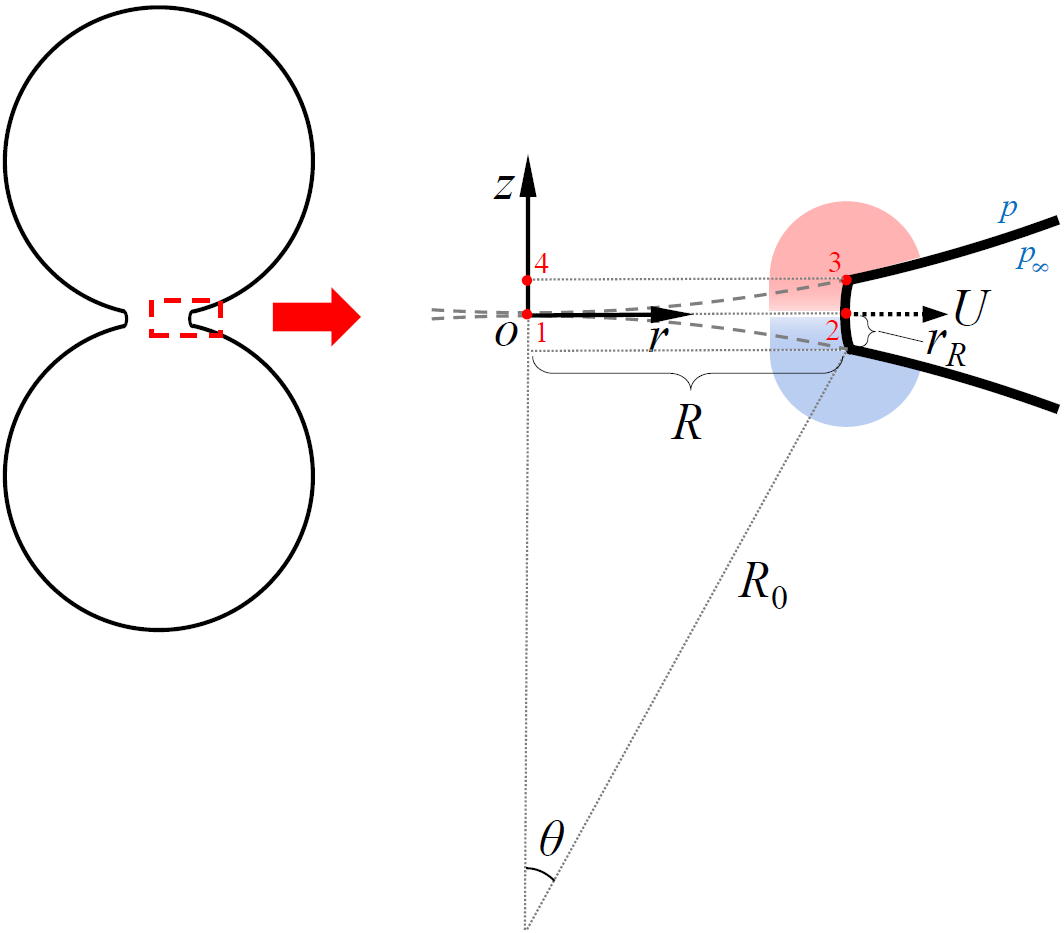}
\caption{A zoomed-in schematic of the neck region between two merging droplets. The red and blue contours illustrate the vorticity distribution localized around the neck.}
\label{fig:1}
\end{center}
\end{figure}

In this letter, we present a theory that significantly rigorizes the previous scaling model \cite{Xia:18a} and contains no empirical constant.
A schematic of the neck between two merging droplets of initial radius $R_0$ is shown in Fig.~\ref{fig:1}. The neck radius, $R$, is defined as the minimum radial distance from the $z$-axis to the neck. Under capillary pressure difference, the neck expands out at a speed of $U(t)$. We assume the flow to be\\ 
(i) \emph{Quasi-steady}, meaning the flow acceleration is mainly associated with the convection induced by the neck movement.\\ 
(ii) \emph{Quasi-radial}, meaning the neck region can be treated as a ring of radius $R$ and width $2r_R$, which is driven by a distributed and quasi-radially directed capillary force \cite{EggersJ:99a}. The capillary force is related to two principle curvatures, $1/R$ and $1/r_R$ \cite{GrossM:13a}, with the latter being the effective curvature in the $zr$-plane.\\ 
(iii) \emph{Localized}, meaning the significant velocity gradients are restricted to the vicinity of the neck as illustrated in Fig.~\ref{fig:1}. This assumption accords with the finding of Paulsen \textit{et al.} \cite{PaulsenJD:11a} that the flow extends over a length comparable to the neck width rather than the neck radius. It follows that the main vortical structure has a length of $O(r_R)$, and the origin (point 1) is considered as the far field where the velocity gradients are effectively zero.\\
(iv) \emph{Geometrical self-similar}, so that the neck width satisfies the simple geometric relation, $r_R/R = \tan{(\theta/2)}$. Under the coalescence regime of $R \ll R_0$, we have $\tan{(\theta/2)} \approx \theta/2 \approx R/(2R_0)$ and, consequently,
\begin{equation}
\label{eq:4}
\frac{r_R}{R} \approx \frac{R}{2R_0} \ll 1,
\end{equation}
which is consistent with other studies \cite{PaulsenJD:11a,GrossM:13a}, but different from Eggers \textit{et al.} \cite{EggersJ:99a} who assumed $r_R/R$ to be higher-order small. This could be responsible for their logarithmic scaling law, which is believed to occur at very early stage of droplet coalescence \cite{Xia:18a}, beyond the resolution of existing experiments.

For the axisymmetric and quasi-steady flow, the $r$-direction N-S equation is expressed as
\begin{equation}
\label{eq:1}
 \rho (u_z \partial_z u_r + u_r \partial_r u_r) = -\partial_r p + \mu \left[\partial_z^2 u_r + \partial_r^2 u_r + \partial_r (\frac{u_r}{r}) \right],
\end{equation}
where $u_z$ and $u_r$ are the velocity components in the $z$- and $r$- directions, respectively, and $p$ is the pressure. Along the $r$-axis, $u_z$ and $\partial_z u_r$ are all zeros owing to the condition of symmetry, so the term $u_z \partial_z u_r$ vanishes in Eq.~\ref{eq:1}. We now integrate Eq.~\ref{eq:1} along the $r$-axis from point 1 ($r=0,z=0$) to point 2 ($r=R,z=0$) as
\begin{equation}
\begin{split}
\label{eq:2}
 &\int_{1\rightarrow2} \left[ \frac{1}{2}\rho\partial_r u^2_r + \partial_r p - \mu \left(\partial_z^2 u_r + \partial_r^2 u_r + \partial_r (\frac{u_r}{r})\right) \right] \mathrm{d}r\\ 
 & = \frac{1}{2}\rho U^2 + p_2-p_1 - \mu \left(\int_{0}^{R} \partial_z^2 u_r \mathrm{d}r + (\partial_r u_r)|_2 + \frac{U}{R}\right)  = 0,
\end{split}
\end{equation}
where the subscripts $_1$ and $_2$ denote the quantities associated with point 1 and 2, respectively. In attaining Eq.~\ref{eq:2}, we have also applied $(u_r)|_1 = 0$ according to the axisymmetric condition, $(\partial_r u_r)|_1 = 0$ following assumption (iii), and $(u_r)|_2 = U(t)$. As the present theory concerns the coalescence of liquid droplets in a gaseous environment, the liquid-gas interface can be treated as a free surface, across which the capillary pressure jump is $p_\infty - p = -2\mu \boldsymbol{n}\cdot \boldsymbol{S} \cdot \boldsymbol{n} + \sigma\kappa$ \citep{TryggvasonG:11a}, where $p_\infty$ is the ambient gas pressure,  $\boldsymbol{n}$ and $\kappa$ are the unit normal vector and curvature of the interface, respectively, and $\boldsymbol{S}$ is the rate-of-strain tensor. Accordingly, the pressures at the far-side droplet and the neck satisfy $p_\infty - p_1 = -2\sigma/R_0$ and $p_\infty - p_2 = -2\mu(\partial_r u_r)|_2+\sigma(1/r_R - 1/R)$, respectively. Here, $p_1$ serves as the pressure at the far-side droplet according to assumption (iii). Subtracting the two equations yields $p_2-p_1 = -\sigma(1/r_R - 1/R + 2/R_0) + 2\mu (\partial_r u_r)|_2$, which can be plugged into Eq.~\ref{eq:2} to obtain 
\begin{equation}
\begin{split}
\label{eq:2-1}
 &\frac{1}{2}\rho U^2 - \sigma \left(\frac{1}{r_R} - \frac{1}{R} + \frac{2}{R_0}\right)\\
 &- \mu \left(\int_{0}^{R} \partial_z^2 u_r \mathrm{d}r  + (\partial_z u_z)|_2 + \frac{2U}{R}\right) = 0.
\end{split}
\end{equation}
Note that the continuity equation, $\partial_z u_z + \partial_r u_r + u_r/r = 0$, has been used in the above derivation. 

The quasi-radial assumption (ii) implies $u_z = 0$ around point 2 and further $(\partial_z u_z)|_2 = 0$ in Eq.~\ref{eq:2-1}. Furthermore, $\partial_z^2 u_r$ can be expressed as
\begin{equation}
\label{eq:2-3}
 \partial_z^2 u_r \approx \frac{(\partial_z u_r)|_{z=r_R}-(\partial_z u_r)|_{z=0}}{r_R} = \frac{(\partial_r u_z + \omega)|_{z=r_R}}{r_R},
\end{equation}
with $(\partial_z u_r)|_{z=0}=0$ by axisymmetry and $\omega = \partial_z u_r - \partial_r u_z$ being the vorticity. Eq.~\ref{eq:2-3} essentially gives a leading-order approximation based on linearizing the strain rate near the plane of symmetry. Integrating Eq.~\ref{eq:2-3} from point 4 ($r=0,z=r_R$) to point 3 ($r=R,z=r_R$) yields
\begin{equation}
\label{eq:2-4}
 \int_{0}^{R} \partial_z^2 u_r \mathrm{d}r \approx \frac{1}{r_R} \left( (u_z)|_{4}^{3} + \int_0^R \omega|_{z=r_R} \mathrm{d}r \right),
\end{equation}
with $(u_z)|_3 = 0$ by assumption (ii) and $(u_z)|_4 = 0$ by assumption (iii). 

\begin{figure}
\begin{center}
\includegraphics[scale=0.63]{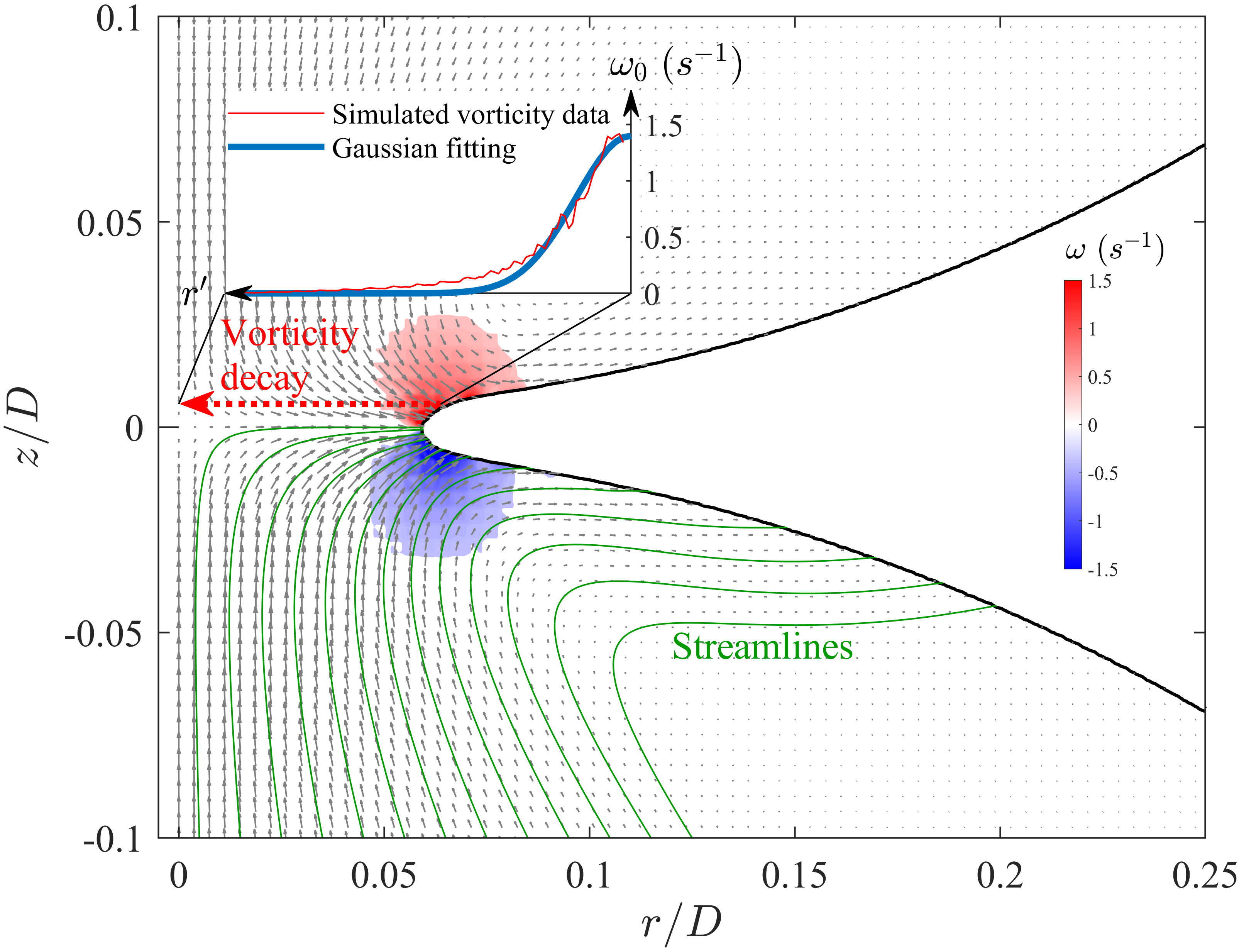}
\caption{The simulated flow field for a representative case of $Oh = 4$. The numerical method is reported in the Supplementary Materials \cite{SM:19a}.}
\label{fig:1-1}
\end{center}
\end{figure}

Lacking \emph{a priori} knowledge of the vorticity field, we seek an approximation of $\omega|_{z=r_R}$ based on the computational observation that in the $orz$ plane the vortex-dynamical effect of the neck movement induces two opposite-sign vortices that are centered around the two edges of the neck, as illustrated in Fig.~\ref{fig:1-1}. This physical picture is also consistent with assumption (iii). The radial vorticity decay displayed in Fig.~\ref{fig:1-1} further implies that the vortex is analogous to a Batchelor vortex \cite{Batchelor:64a} and has a Gaussian vorticity distribution as 
%
\begin{equation}
\label{eq:2-5}
 \omega_0(r') = \frac{U}{r_v}e^{-\left(\frac{r'}{r_v}\right)^2},
\end{equation}
where $r' = R-r$ is the radial location relative to the vortex center located at the neck interface, and $r_v$ is an effective radius of the vortex core. Eq.~\ref{eq:2-5} is also similar to the Oseen-Lamb vortex \cite{WuJZ:06a}, which is an analytical solution to the vorticity diffusion equation.
Recognizing that $\omega_0(r') = \omega(r) = \omega(R-r')$ for $z=r_R$ and $0 \leq r \leq R$, Eq.~\ref{eq:2-4} can be further derived as
\begin{equation}
\label{eq:2-6}
 -\frac{U}{r_R} \int_0^{\frac{R}{r_v}} e^{-\left(\frac{r'}{r_v}\right)^2} \mathrm{d}\left(\frac{r'}{r_v}\right) = -\frac{\sqrt{\pi}U}{2r_R} \mathrm{erf}\left(\frac{R}{r_v}\right) \approx -\frac{\sqrt{\pi}U}{2r_R},
\end{equation}
with $R/r_v \gg 1$ given by assumption (iii) and Eq.~\ref{eq:4}.

We now plug in Eq.~\ref{eq:2-6} to cast Eq.~\ref{eq:2-1} in the form,
\begin{equation}
\label{eq:2-7}
 \frac{1}{2}\rho U^2 - \sigma \left(\frac{1}{r_R} - \frac{1}{R} + \frac{2}{R_0}\right) - \mu \left(- \frac{\sqrt{\pi}U}{2r_R} + \frac{2U}{R}\right) = 0.
\end{equation}
Applying Eq.~\ref{eq:4} and balancing the leading-order terms of Eq.~\ref{eq:2-7} yields 
\begin{equation}
\label{eq:3}
 \rho U^2 - \frac{4\sigma R_0}{R^2} + \frac{2\sqrt{\pi} \mu R_0 U}{R^2} = 0,
\end{equation}
which can be combined with $\dot{R} = \mathrm{d}R/\mathrm{d}t = U$ to derive
\begin{equation}
\label{eq:6}
\frac{\rho \dot{R}^{\ast 2} L^2}{T^2} - \frac{2\sigma D_0}{R^{\ast 2}L^2} + \frac{\sqrt{\pi} \mu D_0 \dot{R}^{\ast}}{R^{\ast 2}LT} = 0,
\end{equation}
where $D_0=2R_0$, $R^{\ast} = R/L$, $\dot{R}^{\ast}=\dot{R}/U$, and $T=L/U$, with $L$, $U$, and $T$ being the characteristic length, velocity, and time scales, respectively.  

The experimental studies of Paulsen \textit{et al.} \cite{PaulsenJD:11a,PaulsenJD:13a} imply the existence of a unified formula for the neck movement given the length and time are scaled properly. Let Eq.~\ref{eq:6} be such a formula, we have 
\begin{equation}
\label{eq:7}
\frac{\rho L^2}{T^2} = \frac{\sigma D_0}{L^2} = \frac{\mu D_0}{LT},
\end{equation}
yielding $L = Oh D_0$ and $T = \mu Oh D_0/\sigma$ where $Oh = \mu/\sqrt{\rho \sigma D_0}$ is the Ohnesorge number. Note that $L$ and $T$ match exactly with the viscous-to-inertial crossover scales found by previous studies \cite{BurtonJC:07a,PaulsenJD:11a,PaulsenJD:13a}. Accordingly, Eq.~\ref{eq:6} takes the dimensionless form,
\begin{equation}
\label{eq:8}
\dot{R}^{\ast 2} - \frac{2}{R^{\ast 2}} + \frac{\sqrt{\pi} \dot{R}^{\ast}}{R^{\ast 2}} = 0.
\end{equation}
We can integrate Eq.~\ref{eq:8} with the initial condition $R^{\ast}(t^{\ast}=0) = 0$, where $t^{\ast} = t/T$, to obtain the exact solution,
\begin{equation}
\begin{split}
\label{eq:9}
t^{\ast} = \frac{\sqrt{\pi} R^{\ast}}{4} + \frac{\sqrt{\pi}}{8}\left[R^{\ast}\sqrt{\frac{8 R^{\ast 2}}{\pi} + 1} + \frac{\sqrt{\pi}}{2\sqrt{2}}\sinh^{-1}\left(\frac{2\sqrt{2} R^{\ast}}{\sqrt{\pi}}\right) \right].
\end{split}
\end{equation}
Eq.~\ref{eq:9} readily dictates the asymptotic behaviors associated with the viscous and inertial regimes. For the inertial regime, $R^{\ast} \gg \sqrt{2\pi}/4$, Eq.~\ref{eq:9} yields
\begin{equation}
\label{eq:10}
t^{\ast} \approx \frac{R^{\ast 2}}{2\sqrt{2}} + O(R^{\ast}).
\end{equation}
For the viscous regimes, $R^{\ast} \ll \sqrt{2\pi}/4$, Eq.~\ref{eq:9} yields 
\begin{equation}
\label{eq:11}
t^{\ast} \approx \frac{\sqrt{\pi}}{4}\left[\frac{3R^{\ast}}{2} + \frac{\sqrt{\pi}}{4\sqrt{2}}\ln\left(\frac{2\sqrt{2} R^{\ast}}{\sqrt{\pi}} + 1\right) \right] \approx \frac{\sqrt{\pi} R^{\ast}}{2} + O(R^{\ast 2}).
\end{equation}
Eq.~\ref{eq:11} can be also reduced to the form of $R \sim t\sigma/\mu$, which is void of any characteristic length. This can be interpreted that the physics of the viscous regime is intermediate self-similar \cite{BarenblattGI:96a}. 
To evaluate our theory, we first write Eq.~\ref{eq:10} in the dimensional form of $R/R_0 \approx c_1(t/\tau_i)^{1/2}$ with $c_1 = 2$, which recovers the $1/2$ power-law scaling for the inertial regime. Similarly, Eq.~\ref{eq:11} can be expressed in the dimensional form of $R/R_0 \approx c_2 t/\tau_v$ with $c_2=2/\sqrt{\pi}$, which gives the linear scaling relation observed from experiments of high-viscosity droplets \cite{AartsDGAL:05a}. As a reference, the fitting coefficients $c_1 = 1.68$ and $c_2 = 1$ were obtained by Paulsen \cite{PaulsenJD:13a} although different values were reported by others \cite{AartsDGAL:05a,Thoroddsen:05a,WuM:04a}. 

\begin{figure}
\begin{center}
\includegraphics[scale=0.37]{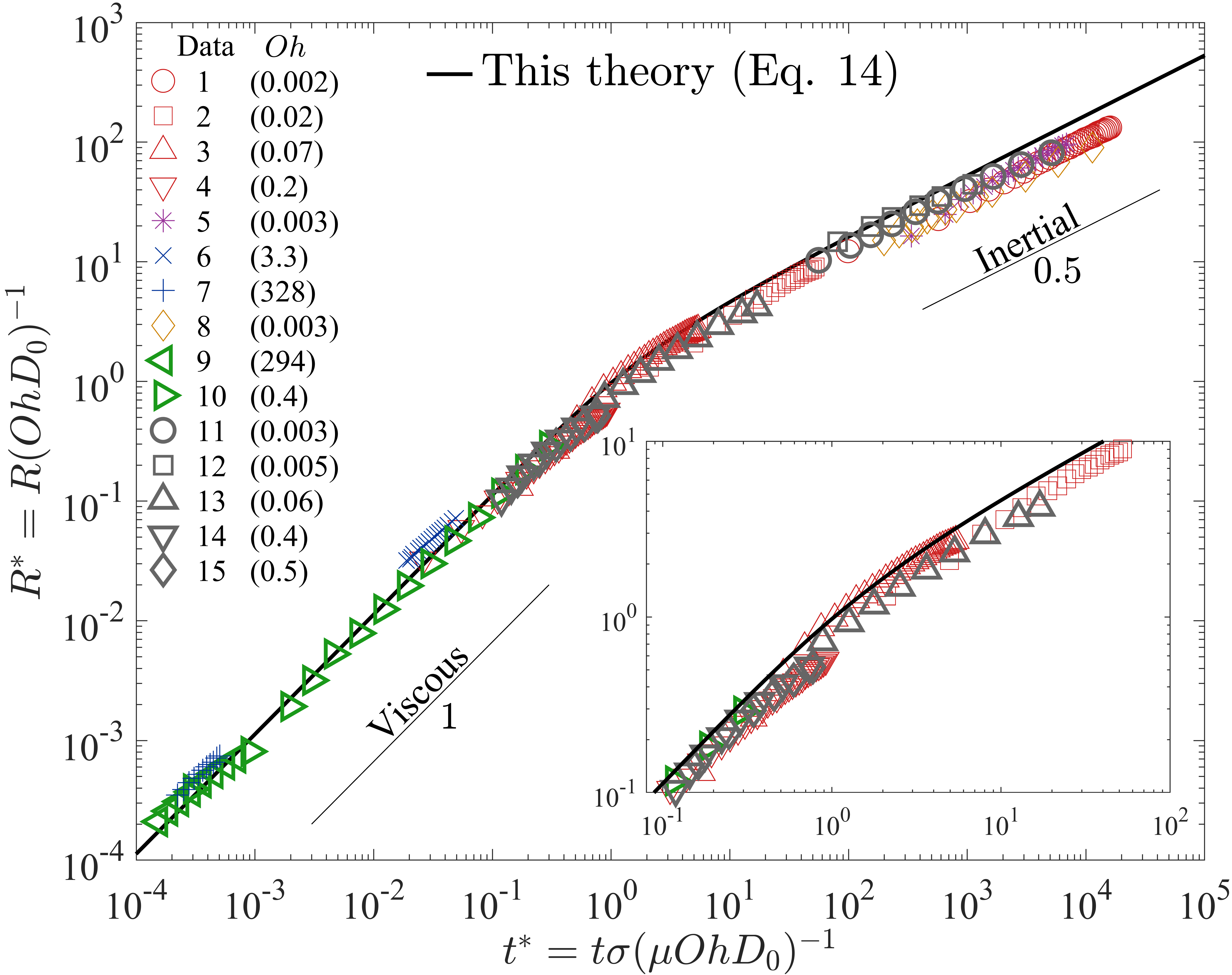}
\caption{Model validation against experimental data from previous studies (see Supplementary Materials \cite{SM:19a} for detailed experimental parameters). A close-up of the crossover regime is shown in the inset plot.}
\label{fig:2}
\end{center}
\end{figure}

Fig.~\ref{fig:2} shows existing experimental data of various $Oh$, corresponding to a variety of fluid types, such as water, silicon oil, and glycerol-salt-water mixture, that are of distinct fluid properties as summarized in the Supplementary Materials \cite{SM:19a}. It is observed that all data tend to collapse onto a single curve, well predicted by the current theory. Considering the assumptions and approximations made in the derivation, the agreement between theory and experiment is quite satisfactory. The theory also captures the asymptotic behaviors of the data in the viscous and inertial regimes. Specifically, the $R^{\ast} \sim t^{\ast}$ and $R^{\ast} \sim \sqrt{t^{\ast}}$ scaling relations show up as $R^{\ast} \ll 1$ and $R^{\ast} \gg 1$, respectively, whereas a clear inflection point can be identified around $R^{\ast} \sim O(1)$ and $t^{\ast} \sim O(1)$, marking the transition from viscous to inertial. It should be emphasized that, although empirical \cite{PaulsenJD:13a} and semi-empirical \cite{Xia:18a} models exist previously, this letter presents the first theory that resolves the unified scaling in the viscous-to-inertial combined coalescence process.  

\begin{figure}
\begin{center}
\includegraphics[scale=0.37]{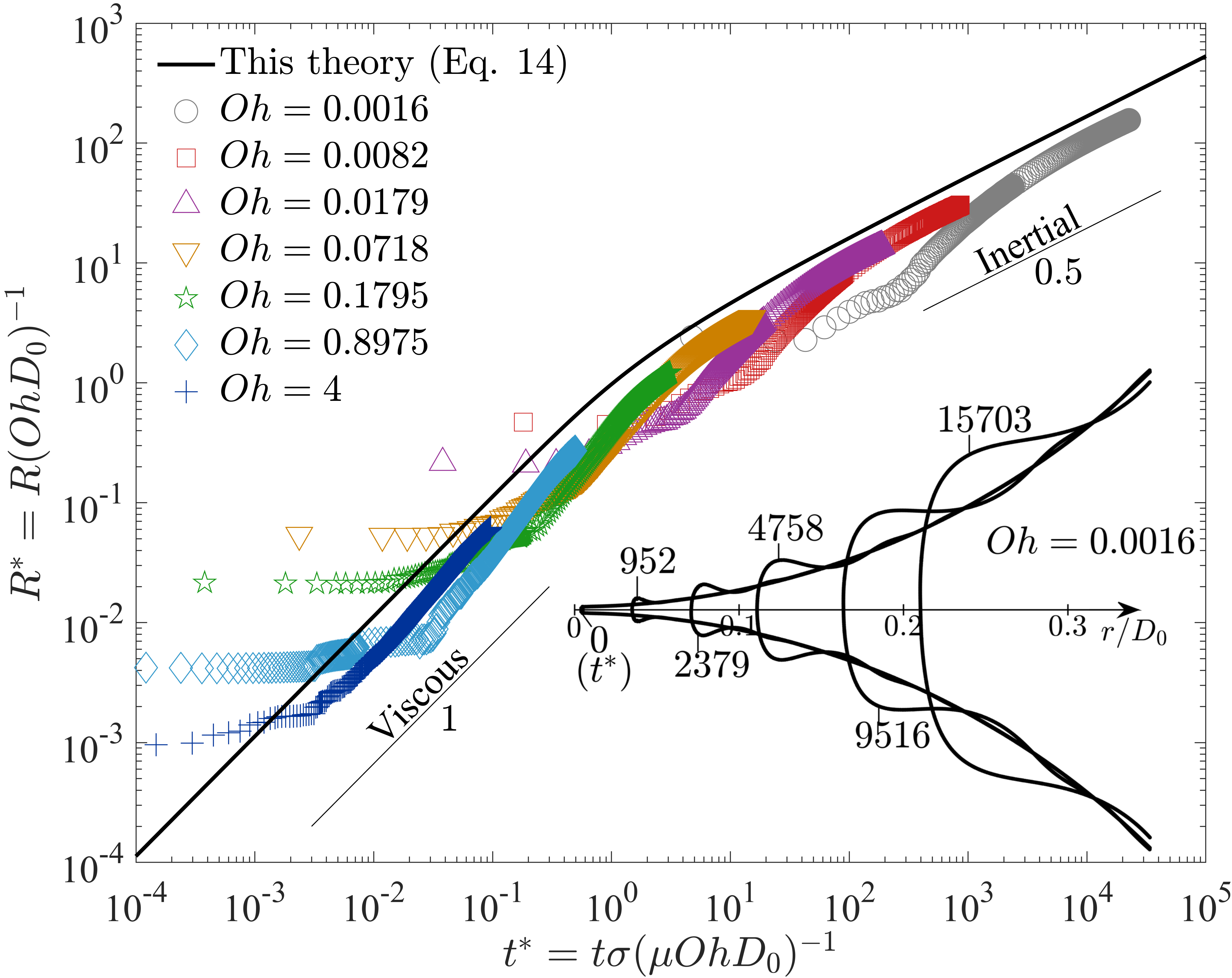}
\caption{Main: validation of the current theory (Eq.~\ref{eq:9}) against simulated neck evolution for droplets of different viscosities ($Oh$). Inset: time evolution of the simulated neck interface for a representative case with $Oh = 0.0016$.}
\label{fig:4}
\end{center}
\end{figure}

Next, we provide further validation of the theory against droplet coalescence simulations of various viscosities. The simulation setup is specified in the Supplementary Materials \cite{SM:19a}. The neck interface evolution for a representative case of $Oh = 0.0016$ is shown in the inset plot of Fig.~\ref{fig:4}. Similar simulations were conducted for $Oh=$ 0.0082, 0.0179, 0.0718, 0.1795, 0.8975, and 4. The corresponding neck radius evolutions are presented in the main plot of Fig.~\ref{fig:4}. It is seen that each simulation data set originates from a finite neck radius, causing the simulated evolution to deviate from the theory. Nevertheless, the later-stage coalescence behavior is less affected by the simulation onset, as each neck evolution curve gradually approaches and then follows its designated scaling, showing that the overall trend of the simulation curves are still captured by the theory. Similar neck evolution behaviors were also observed from previous simulations \cite{SprittlesJE:12a,SprittlesJE:14a}.

Last, this theory suggests that $R^{\ast} = R/(Oh D_0)$ is a criterion segmenting the different coalescence regimes. Although it involves both $R/D_0$ and $Oh$, for different fluids, $Oh$ is the parameter that eventually decides whether the inertial regime could arrive. This is evident from both Fig.~\ref{fig:2} and Fig.~\ref{fig:4} that data in the inertial regime generally corresponds to smaller $Oh$ and vice versa. This criterion has important practical use. For example, Aarts \textit{et al.} \cite{AartsDGAL:05a} considered Data 3 (20 mPa s silicon oil) and Data 4 (50 mPa s silicon oil) to be within the inertial regime, whereas Fig.~\ref{fig:2} clearly shows that Data 3 mainly covers the crossover regime and Data 4 extends from the viscous regime to the crossover regime.


To summarize, this letter presents a theory for the neck evolution during initial coalescence of binary liquid droplets. We have derived and validated a unified solution that applies to the viscous, viscous-to-inertial crossover, and inertial regimes of droplet coalescence. This provides a fundamental framework to support the prominent scaling laws as well as the crossover behaviors observed from previous experimental studies.

We would like to acknowledge the support from the Hong Kong RGC/GRF (PolyU 152217/14E and PolyU 152651/16E) and the ``Open Fund" of State Key Laboratory of Engines (Tianjin University, No. K2018-12).

\bibliography{InitDropCoal_PRL2019_v3}

\newcommand{\AIAAJ}{AIAA J.} \newcommand{\AIAAP}{AIAA Paper}
  \newcommand{\ARMA}{Archive for Rational Mechanics and Analysis}
  \newcommand{\ASMEJFE}{J. Fluids Eng., Trans. ASME} \newcommand{\ASR}{Applied
  Scientific Research} \newcommand{\CF}{Computers Fluids}
  \newcommand{\CJFAS}{Can. J. Fish. Aquat. Sci.}
  \newcommand{\ETFS}{Experimental Thermal and Fluid Science}
  \newcommand{\EF}{Experiments in Fluids} \newcommand{\FDR}{Fluid Dynamics
  Research} \newcommand{\IJHMT}{Int. J. Heat Mass Transfer}
  \newcommand{\JASA}{J. Acoust. Soc. Am.} \newcommand{\JCP}{J. Comp. Physics}
  \newcommand{\JEB}{J. Exp. Biol.} \newcommand{\JFM}{J. Fluid Mech.}
  \newcommand{\JMP}{J. Math. Phys.} \newcommand{\JSC}{J. Scientific Computing}
  \newcommand{\JSP}{J. Stat. Phys.} \newcommand{\JSV}{J. of Sound and
  Vibration} \newcommand{\MC}{Mathematics of Computation}
  \newcommand{\MWR}{Monthly Weather Review} \newcommand{\PAS}{Prog. in
  Aerospace. Sci.} \newcommand{\PCPS}{Proc. Camb. Phil. Soc.}
  \newcommand{\PD}{Physica D} \newcommand{\PRA}{Physical Rev. A}
  \newcommand{\PRE}{Physical Rev. E} \newcommand{\PRL}{Phys. Rev. Lett.}
  \newcommand{\PF}{Phys. Fluids} \newcommand{\PFA}{Phys. Fluids A.}
  \newcommand{\PL}{Phys. Lett.} \newcommand{\PRSLA}{Proc. R. Soc. Lond. A}
  \newcommand{\SIAMJMA}{SIAM J. Math. Anal.} \newcommand{\SIAMJNA}{SIAM J.
  Numer. Anal.} \newcommand{\SIAMJSC}{SIAM J. Sci. Comput.}
  \newcommand{\SIAMJSSC}{SIAM J. Sci. Stat. Comput.}
  \newcommand{\TCFD}{Theoret. Comput. Fluid Dynamics} \newcommand{\ZAMM}{ZAMM}
  \newcommand{\ZAMP}{ZAMP} \newcommand{\ICASER}{ICASE Rep. No.}
  \newcommand{\NASACR}{NASA CR} \newcommand{\NASATM}{NASA TM}
  \newcommand{\NASATP}{NASA TP} \newcommand{\ARFM}{Ann. Rev. Fluid Mech.}
  \newcommand{\WWW}{from {\tt www}.} \newcommand{\CTR}{Center for Turbulence
  Research, Annual Research Briefs} \newcommand{\vonKarman}{von Karman
  Institute for Fluid Dynamics Lecture Series}
\begin{thebibliography}{38}
\expandafter\ifx\csname natexlab\endcsname\relax\def\natexlab#1{#1}\fi
\expandafter\ifx\csname bibnamefont\endcsname\relax
  \def\bibnamefont#1{#1}\fi
\expandafter\ifx\csname bibfnamefont\endcsname\relax
  \def\bibfnamefont#1{#1}\fi
\expandafter\ifx\csname citenamefont\endcsname\relax
  \def\citenamefont#1{#1}\fi
\expandafter\ifx\csname url\endcsname\relax
  \def\url#1{\texttt{#1}}\fi
\expandafter\ifx\csname urlprefix\endcsname\relax\def\urlprefix{URL }\fi
\providecommand{\bibinfo}[2]{#2}
\providecommand{\eprint}[2][]{\url{#2}}

\bibitem[{\citenamefont{Eggers et~al.}(1999)\citenamefont{Eggers, Lister, and
  Stone}}]{EggersJ:99a}
\bibinfo{author}{\bibfnamefont{J.}~\bibnamefont{Eggers}},
  \bibinfo{author}{\bibfnamefont{J.~R.} \bibnamefont{Lister}},
  \bibnamefont{and} \bibinfo{author}{\bibfnamefont{H.~A.} \bibnamefont{Stone}},
  \bibinfo{journal}{J. Fluid Mech.} \textbf{\bibinfo{volume}{401}},
  \bibinfo{pages}{293} (\bibinfo{year}{1999}).

\bibitem[{\citenamefont{Aarts et~al.}(2005)\citenamefont{Aarts, Lekkerkerker,
  Guo, Wegdam, and Bonn}}]{AartsDGAL:05a}
\bibinfo{author}{\bibfnamefont{D.~G. A.~L.} \bibnamefont{Aarts}},
  \bibinfo{author}{\bibfnamefont{H.~N.~W.} \bibnamefont{Lekkerkerker}},
  \bibinfo{author}{\bibfnamefont{H.}~\bibnamefont{Guo}},
  \bibinfo{author}{\bibfnamefont{G.~H.} \bibnamefont{Wegdam}},
  \bibnamefont{and} \bibinfo{author}{\bibfnamefont{D.}~\bibnamefont{Bonn}},
  \bibinfo{journal}{Phys. Rev. Lett.} \textbf{\bibinfo{volume}{95}},
  \bibinfo{pages}{164503} (\bibinfo{year}{2005}).

\bibitem[{\citenamefont{Thoroddsen et~al.}(2005)\citenamefont{Thoroddsen,
  Takehara, and Etoh}}]{Thoroddsen:05a}
\bibinfo{author}{\bibfnamefont{S.~T.} \bibnamefont{Thoroddsen}},
  \bibinfo{author}{\bibfnamefont{K.}~\bibnamefont{Takehara}}, \bibnamefont{and}
  \bibinfo{author}{\bibfnamefont{T.~G.} \bibnamefont{Etoh}},
  \bibinfo{journal}{J. Fluid Mech.} \textbf{\bibinfo{volume}{527}},
  \bibinfo{pages}{85} (\bibinfo{year}{2005}).

\bibitem[{\citenamefont{Paulsen et~al.}(2011)\citenamefont{Paulsen, Burton, and
  Nagel}}]{PaulsenJD:11a}
\bibinfo{author}{\bibfnamefont{J.~D.} \bibnamefont{Paulsen}},
  \bibinfo{author}{\bibfnamefont{J.~C.} \bibnamefont{Burton}},
  \bibnamefont{and} \bibinfo{author}{\bibfnamefont{S.~R.} \bibnamefont{Nagel}},
  \bibinfo{journal}{Phys. Rev. Lett.} \textbf{\bibinfo{volume}{106}},
  \bibinfo{pages}{114501} (\bibinfo{year}{2011}).

\bibitem[{\citenamefont{Xia et~al.}(2017)\citenamefont{Xia, He, Yu, Zhao, and
  Zhang}}]{Xia:17a}
\bibinfo{author}{\bibfnamefont{X.}~\bibnamefont{Xia}},
  \bibinfo{author}{\bibfnamefont{C.}~\bibnamefont{He}},
  \bibinfo{author}{\bibfnamefont{D.}~\bibnamefont{Yu}},
  \bibinfo{author}{\bibfnamefont{J.}~\bibnamefont{Zhao}}, \bibnamefont{and}
  \bibinfo{author}{\bibfnamefont{P.}~\bibnamefont{Zhang}},
  \bibinfo{journal}{Phys. Rev. Fluids} \textbf{\bibinfo{volume}{2}},
  \bibinfo{pages}{113607} (\bibinfo{year}{2017}).

\bibitem[{\citenamefont{Yarin}(2006)}]{YarinAL:06a}
\bibinfo{author}{\bibfnamefont{A.~L.} \bibnamefont{Yarin}},
  \bibinfo{journal}{\ARFM} \textbf{\bibinfo{volume}{38}}, \bibinfo{pages}{159}
  (\bibinfo{year}{2006}).

\bibitem[{\citenamefont{Zhang and Law}(2011)}]{ZhangP:11a}
\bibinfo{author}{\bibfnamefont{P.}~\bibnamefont{Zhang}} \bibnamefont{and}
  \bibinfo{author}{\bibfnamefont{C.~K.} \bibnamefont{Law}},
  \bibinfo{journal}{Phys. Fluids} \textbf{\bibinfo{volume}{23}},
  \bibinfo{pages}{042102} (\bibinfo{year}{2011}).

\bibitem[{\citenamefont{Thoraval et~al.}(2012)\citenamefont{Thoraval, Takehara,
  Etoh, Popinet, Ray, Josserand, Zaleski, and Thoroddsen}}]{ThoravalMJ:12a}
\bibinfo{author}{\bibfnamefont{M.-J.} \bibnamefont{Thoraval}},
  \bibinfo{author}{\bibfnamefont{K.}~\bibnamefont{Takehara}},
  \bibinfo{author}{\bibfnamefont{T.~G.} \bibnamefont{Etoh}},
  \bibinfo{author}{\bibfnamefont{S.}~\bibnamefont{Popinet}},
  \bibinfo{author}{\bibfnamefont{P.}~\bibnamefont{Ray}},
  \bibinfo{author}{\bibfnamefont{C.}~\bibnamefont{Josserand}},
  \bibinfo{author}{\bibfnamefont{S.}~\bibnamefont{Zaleski}}, \bibnamefont{and}
  \bibinfo{author}{\bibfnamefont{S.~T.} \bibnamefont{Thoroddsen}},
  \bibinfo{journal}{Phys. Rev. Lett.} \textbf{\bibinfo{volume}{108}},
  \bibinfo{pages}{264506} (\bibinfo{year}{2012}).

\bibitem[{\citenamefont{Tran et~al.}(2013)\citenamefont{Tran, de~Maleprade,
  Sun, and Lohse}}]{TranT:13a}
\bibinfo{author}{\bibfnamefont{T.}~\bibnamefont{Tran}},
  \bibinfo{author}{\bibfnamefont{H.}~\bibnamefont{de~Maleprade}},
  \bibinfo{author}{\bibfnamefont{C.}~\bibnamefont{Sun}}, \bibnamefont{and}
  \bibinfo{author}{\bibfnamefont{D.}~\bibnamefont{Lohse}}, \bibinfo{journal}{J.
  Fluid Mech.} \textbf{\bibinfo{volume}{726}}, \bibinfo{pages}{R3}
  (\bibinfo{year}{2013}).

\bibitem[{\citenamefont{Kavehpour}(2015)}]{KavehpourHP:15a}
\bibinfo{author}{\bibfnamefont{H.~P.} \bibnamefont{Kavehpour}},
  \bibinfo{journal}{\ARFM} \textbf{\bibinfo{volume}{47}}, \bibinfo{pages}{245}
  (\bibinfo{year}{2015}).

\bibitem[{\citenamefont{Tang et~al.}(2016)\citenamefont{Tang, Zhao, Zhang, Law,
  and Huang}}]{ZhangP:16a}
\bibinfo{author}{\bibfnamefont{C.}~\bibnamefont{Tang}},
  \bibinfo{author}{\bibfnamefont{J.}~\bibnamefont{Zhao}},
  \bibinfo{author}{\bibfnamefont{P.}~\bibnamefont{Zhang}},
  \bibinfo{author}{\bibfnamefont{C.~K.} \bibnamefont{Law}}, \bibnamefont{and}
  \bibinfo{author}{\bibfnamefont{Z.}~\bibnamefont{Huang}}, \bibinfo{journal}{J.
  Fluid Mech.} \textbf{\bibinfo{volume}{795}}, \bibinfo{pages}{671}
  (\bibinfo{year}{2016}).

\bibitem[{\citenamefont{He et~al.}(2019)\citenamefont{He, Xia, and
  Zhang}}]{Xia:19a}
\bibinfo{author}{\bibfnamefont{C.}~\bibnamefont{He}},
  \bibinfo{author}{\bibfnamefont{X.}~\bibnamefont{Xia}}, \bibnamefont{and}
  \bibinfo{author}{\bibfnamefont{P.}~\bibnamefont{Zhang}},
  \bibinfo{journal}{Phys. Fluids} \textbf{\bibinfo{volume}{31}},
  \bibinfo{pages}{052004} (\bibinfo{year}{2019}).

\bibitem[{\citenamefont{Frenkel}(1945)}]{FrenkelJJ:45a}
\bibinfo{author}{\bibfnamefont{J.~J.} \bibnamefont{Frenkel}},
  \bibinfo{journal}{J. Phys.} \textbf{\bibinfo{volume}{9}},
  \bibinfo{pages}{385} (\bibinfo{year}{1945}).

\bibitem[{\citenamefont{Hopper}(1993)}]{HopperRW:93a}
\bibinfo{author}{\bibfnamefont{R.~W.} \bibnamefont{Hopper}},
  \bibinfo{journal}{J. Am. Ceram. Soc.} \textbf{\bibinfo{volume}{76}},
  \bibinfo{pages}{2947} (\bibinfo{year}{1993}).

\bibitem[{\citenamefont{Hopper}(1990)}]{HopperRW:90a}
\bibinfo{author}{\bibfnamefont{R.~W.} \bibnamefont{Hopper}},
  \bibinfo{journal}{J. Fluid Mech.} \textbf{\bibinfo{volume}{213}},
  \bibinfo{pages}{349} (\bibinfo{year}{1990}).

\bibitem[{\citenamefont{Hopper}(1992)}]{HopperRW:92a}
\bibinfo{author}{\bibfnamefont{R.~W.} \bibnamefont{Hopper}},
  \bibinfo{journal}{J. Fluid Mech.} \textbf{\bibinfo{volume}{243}},
  \bibinfo{pages}{171} (\bibinfo{year}{1992}).

\bibitem[{\citenamefont{Duchemin et~al.}(2003)\citenamefont{Duchemin, Eggers,
  and Josseran}}]{DucheminL:03a}
\bibinfo{author}{\bibfnamefont{L.}~\bibnamefont{Duchemin}},
  \bibinfo{author}{\bibfnamefont{J.}~\bibnamefont{Eggers}}, \bibnamefont{and}
  \bibinfo{author}{\bibfnamefont{C.}~\bibnamefont{Josseran}},
  \bibinfo{journal}{J. Fluid Mech.} \textbf{\bibinfo{volume}{487}},
  \bibinfo{pages}{167} (\bibinfo{year}{2003}).

\bibitem[{\citenamefont{Yao et~al.}(2005)\citenamefont{Yao, Maris, Pennington,
  and Seidel}}]{YaoW:05a}
\bibinfo{author}{\bibfnamefont{W.}~\bibnamefont{Yao}},
  \bibinfo{author}{\bibfnamefont{H.~J.} \bibnamefont{Maris}},
  \bibinfo{author}{\bibfnamefont{P.}~\bibnamefont{Pennington}},
  \bibnamefont{and} \bibinfo{author}{\bibfnamefont{G.~M.}
  \bibnamefont{Seidel}}, \bibinfo{journal}{Phys. Rev. E}
  \textbf{\bibinfo{volume}{71}}, \bibinfo{pages}{016309}
  (\bibinfo{year}{2005}).

\bibitem[{\citenamefont{Case and Nagel}(2008)}]{CaseSC:08a}
\bibinfo{author}{\bibfnamefont{S.~C.} \bibnamefont{Case}} \bibnamefont{and}
  \bibinfo{author}{\bibfnamefont{S.~R.} \bibnamefont{Nagel}},
  \bibinfo{journal}{Phys. Rev. Lett.} \textbf{\bibinfo{volume}{100}},
  \bibinfo{pages}{084503} (\bibinfo{year}{2008}).

\bibitem[{\citenamefont{Fezzaa and Wang}(2008)}]{FezzaaK:08a}
\bibinfo{author}{\bibfnamefont{K.}~\bibnamefont{Fezzaa}} \bibnamefont{and}
  \bibinfo{author}{\bibfnamefont{Y.}~\bibnamefont{Wang}},
  \bibinfo{journal}{Phys. Rev. Lett.} \textbf{\bibinfo{volume}{100}},
  \bibinfo{pages}{104501} (\bibinfo{year}{2008}).

\bibitem[{\citenamefont{Sprittles and Shikhmurzaev}(2012)}]{SprittlesJE:12a}
\bibinfo{author}{\bibfnamefont{J.~E.} \bibnamefont{Sprittles}}
  \bibnamefont{and} \bibinfo{author}{\bibfnamefont{Y.~D.}
  \bibnamefont{Shikhmurzaev}}, \bibinfo{journal}{Phys. Fluids}
  \textbf{\bibinfo{volume}{24}}, \bibinfo{pages}{122105}
  (\bibinfo{year}{2012}).

\bibitem[{\citenamefont{Sprittles and Shikhmurzaev}(2014)}]{SprittlesJE:14a}
\bibinfo{author}{\bibfnamefont{J.~E.} \bibnamefont{Sprittles}}
  \bibnamefont{and} \bibinfo{author}{\bibfnamefont{Y.~D.}
  \bibnamefont{Shikhmurzaev}}, \bibinfo{journal}{J. Fluid Mech.}
  \textbf{\bibinfo{volume}{751}}, \bibinfo{pages}{480} (\bibinfo{year}{2014}).

\bibitem[{\citenamefont{Wu et~al.}(2004)\citenamefont{Wu, Cubaud, and
  Ho}}]{WuM:04a}
\bibinfo{author}{\bibfnamefont{M.}~\bibnamefont{Wu}},
  \bibinfo{author}{\bibfnamefont{T.}~\bibnamefont{Cubaud}}, \bibnamefont{and}
  \bibinfo{author}{\bibfnamefont{C.-M.} \bibnamefont{Ho}},
  \bibinfo{journal}{Phys. Fluids} \textbf{\bibinfo{volume}{16}},
  \bibinfo{pages}{L51} (\bibinfo{year}{2004}).

\bibitem[{\citenamefont{Burton and Taborek}(2007)}]{BurtonJC:07a}
\bibinfo{author}{\bibfnamefont{J.~C.} \bibnamefont{Burton}} \bibnamefont{and}
  \bibinfo{author}{\bibfnamefont{P.}~\bibnamefont{Taborek}},
  \bibinfo{journal}{Phys. Rev. Lett.} \textbf{\bibinfo{volume}{98}},
  \bibinfo{pages}{224502} (\bibinfo{year}{2007}).

\bibitem[{\citenamefont{Case}(2009)}]{CaseSC:09a}
\bibinfo{author}{\bibfnamefont{S.~C.} \bibnamefont{Case}},
  \bibinfo{journal}{Phys. Rev. E} \textbf{\bibinfo{volume}{79}},
  \bibinfo{pages}{026307} (\bibinfo{year}{2009}).

\bibitem[{\citenamefont{Eiswirth et~al.}(2012)\citenamefont{Eiswirth, Bart,
  Ganguli, and Kenig}}]{EiswirthRT:12a}
\bibinfo{author}{\bibfnamefont{R.~T.} \bibnamefont{Eiswirth}},
  \bibinfo{author}{\bibfnamefont{H.~J.} \bibnamefont{Bart}},
  \bibinfo{author}{\bibfnamefont{A.~A.} \bibnamefont{Ganguli}},
  \bibnamefont{and} \bibinfo{author}{\bibfnamefont{E.~Y.} \bibnamefont{Kenig}},
  \bibinfo{journal}{Phys. Fluids} \textbf{\bibinfo{volume}{24}},
  \bibinfo{pages}{062108} (\bibinfo{year}{2012}).

\bibitem[{\citenamefont{Pothier and Lewis}(2012)}]{PothierJC:12a}
\bibinfo{author}{\bibfnamefont{J.~C.} \bibnamefont{Pothier}} \bibnamefont{and}
  \bibinfo{author}{\bibfnamefont{L.~J.} \bibnamefont{Lewis}},
  \bibinfo{journal}{Phys. Rev. B} \textbf{\bibinfo{volume}{85}},
  \bibinfo{pages}{115447} (\bibinfo{year}{2012}).

\bibitem[{\citenamefont{Gross et~al.}(2013)\citenamefont{Gross, Steinbach,
  Raabe, and Varnik}}]{GrossM:13a}
\bibinfo{author}{\bibfnamefont{M.}~\bibnamefont{Gross}},
  \bibinfo{author}{\bibfnamefont{I.}~\bibnamefont{Steinbach}},
  \bibinfo{author}{\bibfnamefont{D.}~\bibnamefont{Raabe}}, \bibnamefont{and}
  \bibinfo{author}{\bibfnamefont{F.}~\bibnamefont{Varnik}},
  \bibinfo{journal}{Phys. Fluids} \textbf{\bibinfo{volume}{25}},
  \bibinfo{pages}{052101} (\bibinfo{year}{2013}).

\bibitem[{\citenamefont{Ristenpart et~al.}(2006)\citenamefont{Ristenpart,
  McCalla, Roy, and Stone}}]{StoneHA:06a}
\bibinfo{author}{\bibfnamefont{W.~D.} \bibnamefont{Ristenpart}},
  \bibinfo{author}{\bibfnamefont{P.~M.} \bibnamefont{McCalla}},
  \bibinfo{author}{\bibfnamefont{R.~V.} \bibnamefont{Roy}}, \bibnamefont{and}
  \bibinfo{author}{\bibfnamefont{H.~A.} \bibnamefont{Stone}},
  \bibinfo{journal}{Phys. Rev. Lett.} \textbf{\bibinfo{volume}{97}},
  \bibinfo{pages}{064501} (\bibinfo{year}{2006}).

\bibitem[{\citenamefont{Hern\'{a}ndez-S\'{a}nchez
  et~al.}(2012)\citenamefont{Hern\'{a}ndez-S\'{a}nchez, Lubbers, Eddi, and
  Snoeijer}}]{EddiA:12a}
\bibinfo{author}{\bibfnamefont{J.~F.} \bibnamefont{Hern\'{a}ndez-S\'{a}nchez}},
  \bibinfo{author}{\bibfnamefont{L.~A.} \bibnamefont{Lubbers}},
  \bibinfo{author}{\bibfnamefont{A.}~\bibnamefont{Eddi}}, \bibnamefont{and}
  \bibinfo{author}{\bibfnamefont{J.~H.} \bibnamefont{Snoeijer}},
  \bibinfo{journal}{Phys. Rev. Lett.} \textbf{\bibinfo{volume}{108}},
  \bibinfo{pages}{184502} (\bibinfo{year}{2012}).

\bibitem[{\citenamefont{Eddi et~al.}(2013)\citenamefont{Eddi, Winkels, and
  Snoeijer}}]{EddiA:13a}
\bibinfo{author}{\bibfnamefont{A.}~\bibnamefont{Eddi}},
  \bibinfo{author}{\bibfnamefont{K.~G.} \bibnamefont{Winkels}},
  \bibnamefont{and} \bibinfo{author}{\bibfnamefont{J.~H.}
  \bibnamefont{Snoeijer}}, \bibinfo{journal}{Phys. Rev. Lett.}
  \textbf{\bibinfo{volume}{111}}, \bibinfo{pages}{144502}
  (\bibinfo{year}{2013}).

\bibitem[{\citenamefont{Paulsen}(2013)}]{PaulsenJD:13a}
\bibinfo{author}{\bibfnamefont{J.~D.} \bibnamefont{Paulsen}},
  \bibinfo{journal}{Phys. Rev. E} \textbf{\bibinfo{volume}{88}},
  \bibinfo{pages}{063010} (\bibinfo{year}{2013}).

\bibitem[{\citenamefont{Xia et~al.}(2018)\citenamefont{Xia, He, and
  Zhang}}]{Xia:18a}
\bibinfo{author}{\bibfnamefont{X.}~\bibnamefont{Xia}},
  \bibinfo{author}{\bibfnamefont{C.}~\bibnamefont{He}}, \bibnamefont{and}
  \bibinfo{author}{\bibfnamefont{P.}~\bibnamefont{Zhang}},
  \bibinfo{journal}{arXiv preprint arXiv:1803.05789}  (\bibinfo{year}{2018}).

\bibitem[{\citenamefont{Tryggvason et~al.}(2011)\citenamefont{Tryggvason,
  Scardovelli, and Zaleski}}]{TryggvasonG:11a}
\bibinfo{author}{\bibfnamefont{G.}~\bibnamefont{Tryggvason}},
  \bibinfo{author}{\bibfnamefont{R.}~\bibnamefont{Scardovelli}},
  \bibnamefont{and} \bibinfo{author}{\bibfnamefont{S.}~\bibnamefont{Zaleski}},
  \emph{\bibinfo{title}{Direct Numerical Simulations of Gas-Liquid Multiphase
  Flows}} (\bibinfo{publisher}{Cambridge University Press},
  \bibinfo{address}{Cambridge, UK}, \bibinfo{year}{2011}).

\bibitem[{\citenamefont{Batchelor}(1964)}]{Batchelor:64a}
\bibinfo{author}{\bibfnamefont{G.~K.} \bibnamefont{Batchelor}},
  \bibinfo{journal}{J. Fluid Mech.} \textbf{\bibinfo{volume}{20}},
  \bibinfo{pages}{645} (\bibinfo{year}{1964}).

\bibitem[{\citenamefont{Wu et~al.}(2006)\citenamefont{Wu, Ma, and
  Zhou}}]{WuJZ:06a}
\bibinfo{author}{\bibfnamefont{J.~Z.} \bibnamefont{Wu}},
  \bibinfo{author}{\bibfnamefont{H.~Y.} \bibnamefont{Ma}}, \bibnamefont{and}
  \bibinfo{author}{\bibfnamefont{M.~D.} \bibnamefont{Zhou}},
  \emph{\bibinfo{title}{Vorticity and Vortex Dynamics}}
  (\bibinfo{publisher}{Springer}, \bibinfo{year}{2006}).

\bibitem[{\citenamefont{{See Supplementary Materials for more numerical details
  and a parameter list of the experimental data in Fig. 3.}}()}]{SM:19a}
\bibinfo{author}{\bibnamefont{{See Supplementary Materials for more numerical
  details and a parameter list of the experimental data in Fig. 3.}}}

\bibitem[{\citenamefont{Barenblatt}(1996)}]{BarenblattGI:96a}
\bibinfo{author}{\bibfnamefont{G.~I.} \bibnamefont{Barenblatt}},
  \emph{\bibinfo{title}{Scaling, self-similarity, and intermediate
  asymptotics}} (\bibinfo{publisher}{Cambridge University Press},
  \bibinfo{address}{Cambridge, UK}, \bibinfo{year}{1996}).

\end{thebibliography}


\begin{thebibliography}{99}

  \bibitem{Yarin2006}
A.L.~Yarin, Annu. Rev. Fluid Mech. {\bf 38}, 159  (2006).

  \bibitem{Gorokhovski2008}
M.~Gorokhovski and M.~Herrmann, Annu. Rev. Fluid Mech. {\bf 40}, 343 (2008).

  \bibitem{Kavehpour2015}
H.P.~Kavehpour, Annu. Rev. Fluid Mech. {\bf 47}, 245 (2015).

  \bibitem{Low1982}
T.B.~Low and R.~List, J. Atmos. Sci. {\bf 39}, 1591 (1982).

  \bibitem{Chen2007}
R.-H.~Chen, Appl. Therm. Eng. {\bf 27}, 604 (2007). 

  \bibitem{Derby2010}
B.~Derby, Annu. Rev. Mater. Res. {\bf 40}, 395 (2010).

  \bibitem{Tang2016}
C.~Tang, J.~Zhao, P.~Zhang, C.K.~Law, and Z.~Huang, J. Fluid Mech. {\bf 795}, 671 (2016).

  \bibitem{Xia2017}
X.~Xia, C.~He, D.~Yu, J.~Zhao, and P.~Zhang, Phys. Rev. Fluids {\bf 2}, 113607 (2017). 

  \bibitem{vandeVorst1994}
G.A.L.~van de Vorst, Technische Universiteit Eindhoven (1994). 

  \bibitem{Dreher1999}
T.M.~Dreher, J.~Glass, A.J.~O'Connor, and G.W.~Stevens, AIChE J. {\bf 45}, 1182 (1999). 

  \bibitem{Squires2005}
T.M.~Squires and S.R.~Quake, Rev. Mod. Phys. {\bf 77}, 977 (2005).

  \bibitem{Frenkel1945}
J.~Frenkel, J. Phys. (Moscow) {\bf 9}, 385 (1945).

  \bibitem{Hopper1993a}
R.W.~Hopper, J. Am. Ceram. Soc. {\bf 76}, 2947 (1993).

  \bibitem{Hopper1990}
R.W.~Hopper, J. Fluid Mech. {\bf 213}, 349 (1990).

  \bibitem{Hopper1993b}
R.W.~Hopper, J. Am. Ceram. Soc. {\bf 76}, 2953 (1993). 

  \bibitem{Eggers1999}
J.~Eggers, J.R.~Lister, and H.A.~Stone, J. Fluid Mech. {\bf 401}, 293 (1999). 

  \bibitem{Duchemin2003}
L. Duchemin, J. Eggers, and C. Josserand, J. Fluid Mech. {\bf 487}, 167 (2003).

  \bibitem{Aarts2005}
D.G.A.L.~Aarts, H.N.W.~Lekkerkerker, H.~Guo, G.H.~Wegdam, and D.~Bonn, Phys. Rev. Lett. {\bf 95}, 164503 (2005).

  \bibitem{Thoroddsen2005}
S.T.~Thoroddsen, K.~Takehara, and T.G.~Etoh, J. Fluid Mech. {\bf 527}, 85 (2005).

  \bibitem{Yao2005}
W.~Yao, H.J.~Maris, P.~Pennington, and G.M.~Seidel, Phys. Rev. E {\bf 71}, 016309 (2005).

  \bibitem{Case2008}
S.C.~Case and S.R.~Nagel, Phys. Rev. Lett. {\bf 100}, 084503 (2008).

  \bibitem{Fezzaa2008}
K.~Fezzaa and Y.~Wang, Phys. Rev. lett. {\bf 100}, 104501 (2008).

  \bibitem{Paulsen2011}
J.D.~Paulsen, J.C.~Burton, and S.R.~Nagel, Phys. Rev. Lett. {\bf 106}, 114501 (2011).

  \bibitem{Wu2004}
M.~Wu, T.~Cubaud, and C.-M.~Ho, Phys. Fluids {\bf 16}, L51 (2004).

  \bibitem{Burton2007}
J.C.~Burton and P.~Taborek, Phys. Rev. Lett. {\bf 98}, 224502 (2007).

  \bibitem{Case2009}
S.C.~Case, Phys. Rev. E {\bf 79}, 026307 (2009).

  \bibitem{Eiswirth2012}
R.T.~Eiswirth, H.J.~Bart, A.A.~Ganguli, and E.Y.~Kenig, Phys. Fluids {\bf 24}, 062108 (2012).

  \bibitem{Pothier2012}
J.C.~Pothier and L.J.~Lewis, Phys. Rev. B {\bf 85}, 115447 (2012).

  \bibitem{Gross2013}
M. Gross, I. Steinbach, D. Raabe, and F. Varnik, Phys. Fluids {\bf 25}, 052101 (2013).

  \bibitem{Sprittles2012}
J.E.~Sprittles and Y.D.~Shikhmurzaev, Phys. Fluids {\bf 24}, 122105 (2012).

  \bibitem{Ristenpart2006}
W.D.~Ristenpart, P.M.~McCalla, R.V.~Roy, and H.A.~Stone, Phys. Rev. Lett. {\bf 97}, 064501 (2006).

  \bibitem{Sanchez2012}
J.F.~Hern\'{a}ndez-S\'{a}nchez, L.A.~Lubbers, A.~Eddi, and J.H.~Snoeijer, Phys. Rev. Lett. {\bf 109}, 184502 (2012).

  \bibitem{Eddi2013}
A.~Eddi, K.G.~Winkels, and J.H.~Snoeijer, Phys. Rev. Lett. {\bf 111}, 144502 (2013).

  \bibitem{Paulsen2013}
J.D.~Paulsen, Phys. Rev. E {\bf 88}, 063010 (2013).

  \bibitem{Tryggvason2011}
G.~Tryggvason, R.~Scardovelli, and S.~Zaleski, \textit{Direct Numerical Simulations of Gas-Liquid Multiphase Flows} (Cambridge University Press, Cambridge, 2011).

  \bibitem{Barenblatt1996}
G.I.~Barenblatt, \textit{Scaling, self-similarity, and intermediate asymptotics} (Cambridge University Press, Cambridge, 1996).

  \bibitem{Paulsen2012}
J.D.~Paulsen, J.C.~Burton, S.R.~Nagel, S.~Appathurai, M.T.~Harris, and O.A.~Basaran, Proc. Natl. Acad. Sci. USA {\bf 109}, 6857 (2012).

  \bibitem{SM}
See Supplemental Materials for a detailed derivation of the velocity integral in \cite{Eggers1999} and relevant parameters of previous experimental data.

  \bibitem{Scardovelli1999}
R.~Scardovelli and S.~Zaleski, Annu. Rev. Fluid Mech. {\bf 31}, 567 (1999).

  \bibitem{Tryggvason2006}
G.~Tryggvason, A.~Esmaeeli, J.~Lu, and S.~Biswas, Fluid Dyn. Res. {\bf 38}, 660 (2006).

  \bibitem{Popinet2003}
S.~Popinet, J. Comput. Phys. {\bf 190}, 572 (2003).

  \bibitem{Popinet2009}
S.~Popinet, J. Comput. Phys. {\bf 228}, 5838 (2009).

  \bibitem{Chen2014a}
X.~Chen and V.~Yang, J. Comput. Phys. {\bf 269}, 22 (2014).

  \bibitem{Chen2014b}
X.~Chen and V.~Yang, Phys. Fluids {\bf 26}, 102104 (2014).

  \bibitem{Agbaglah2015}
G.~Agbaglah, M.-J.~Thoraval, S.T.~Thoroddsen, L.V.~Zhang, K.~Fezzaa, and R.D.~Deegan, J. Fluid Mech. {\bf 764}, R1 (2015).

  \bibitem{Thoraval2016}
M.-J.~Thoraval, Y.~Li, and S.T.~Thoroddsen, Phys. Rev. E {\bf 93}, 033128 (2016).

  \bibitem{Sprittles2014}
J.E.~Sprittles and Y.D.~Shikhmurzaev, J. Fluid Mech. {\bf 751}, 480 (2014).

\end{thebibliography}

\end{document}